\documentclass[preprint,12pt]{elsarticle}

\usepackage{graphicx}
\usepackage{amssymb}
\usepackage{amsmath}

\usepackage{lineno}
\usepackage{textcomp}

\journal{Composite Structures}

\begin{document}

\begin{frontmatter}


\title{Polystyrene-based nanocomposites with different fillers: fabrication and mechanical properties}



\author[label1]{O.A. Moskalyuk}
\author[label2]{A.V. Belashov}
\author[label2]{Y.M. Beltukov}
\author[label3]{E.M. Ivan'kova}
\author[label3]{E.N. Popova}
\author[label2]{I.V. Semenova \corref{cor1}}
\author[label3]{V.Y. Yelokhovsky}
\author[label3]{V.E. Yudin}

\address[label1]{St.Petersburg State University of Industrial Technologies and Design, 18, Bolshaya Morskaya str., St. Petersburg, 191186, Russia}
\address[label2]{Ioffe Institute, 26 Polytekhnicheskaya, St.Petersburg, 194021, Russia}
\address[label3]{Institute of Macromolecular Compounds, 31, Bolshoy pr. V.O., St. Petersburg, 199004, Russia}

\cortext[cor1]{irina.semenova@mail.ioffe.ru}

\begin{abstract}
The paper presents a comprehensive  analysis  of  elastic  properties of polystyrene-based nanocomposites  filled  with  different  types  of inclusions: small spherical particles ($SiO_{2}$ and  $Al_{2}O_{3}$),  alumosilicates (montmorillonite,  halloysite natural tubules  and  Mica) and  carbon  nanofillers  (carbon black  and  multi-walled carbon nanotubes). 
Composites were fabricated by melt technology.  
The analysis of composite melts showed that the introduction of Montmorillonite, Multi-walled carbon nanotubes, and $Al_{2}O_{3}$ particles provided an increase in melt viscosity by an average of 2 to 5 orders of magnitude over the pure polystyrene.
Block samples of composites with different filler concentrations were prepared, and their linear and nonlinear elastic properties were studied. 
The introduction of more rigid particles led to a more profound increase in the elastic modulus of the composite, with the highest rise of about 80\% obtained with carbon fillers. Carbon black particles provided also an enhanced strength at break of about 20\% higher than that of pure polystyrene.  

The nonlinear elastic moduli of composites were shown to be more sensitive to addition of filler particles to the polymer matrix than the linear ones.  The nonlinearity coefficient $\beta$ comprising the combination of linear and nonlinear elastic moduli of a material demonstrated considerable changes correlating with changes of the Young's modulus. The absolute value of $\beta$ showed rise in 1.5--1.6 times in the CB- and HNT-containing composites as compared to that of pure PS.
The changes in nonlinear elasticity of fabricated composites were compared with measurements of the parameters of bulk nonlinear strain waves in them. Variations of wave velocity and decay decrement correlated with observed enhancement of materials nonlinearity.

\end{abstract}


\end{frontmatter}


\section{Introduction}
\label{S:1}

Micro- and nanostructured  composites become very popular nowadays in various engineering applications, see e.g.~\cite{multifunc2016}.
Most of them contain ordered inclusions, i.e., they are manufactured in the form of a matrix filled with oriented filaments made of another material. Numerous examples can be mentioned in this context, from rebar-reinforced concrete to carbon- and boron-reinforced composites for solar batteries. Disordered composites are of common interest as well, 
with most intriguing ones being filled with nanoparticles, which can drastically improve a resulting material strength, stiffness and other physical parameters providing at the same time a uniform material at the macroscale. 
Numerous examples were published recently demonstrating significant advantages of nanocomposites over matrix materials, see e.g.~\cite{gibson,Optimiz_2010,Koratkara2005,Uddin2008,bhattacharya2016,Graphene2017,review-carbon2019}, in particular in view of their enhanced mechanical properties and potential multifunctionality. 
Composites on the base of polymer matrices with various nano-sized fillers is one of the most widely used classes of nanocomposites. 


However, in view of mechanical properties a major disadvantage, common for all polymer nanocomposites, is low predictability of resultant parameters. The uncertainty is mainly due to nonuniform distribution and agglomeration of filler particles in a matrix causing areas with reduced interfacial interaction between the matrix and the filler. The existing experimental data demonstrate considerable variations of measured parameters for the same compositions of materials \cite{review-carbon2019}. 

In the first approximation, elastic properties of a material are governed by the linear elastic moduli, which characterize elastic stresses of a solid at small strains. Since linear moduli give a quadratic contribution to the elastic energy, they are also called the second-order moduli. Linear elastic properties of composite materials have been extensively studied both in theory and experiments and expressions for second-order moduli of elasticity have been derived in various models, see e.g.~\cite{eshelby1957,halpin-kardos76}. It is worth noting though that at comparatively low volume concentrations (up to 10--15\%) of a filler the effective moduli of a composite calculated in terms of the above mentioned models give close values.
However, theoretical estimations of elastic moduli values are, as a rule, much higher than the results of measurements. It may be caused by an assumption of regular (uniform) distribution of nanoinclusions in a polymer, that is far from a real material structure due to manufacturing approaches.

Inhomogeneous distribution of nanoinclusions in a matrix and their micron-sized agglomerates lead to inhomogenenous strength in the composite volume and appearance of local finite-size areas with lower stiffness. For this reason the micromechanical models were refined to take into consideration an intrinsic small-scale inhomogeneity of nanocomposites, and its impact on the structure strength, see, e.g.~\cite{matveeva11,mori-tanaka,ji-cao-feng}. 

At higher strains nonlinear elasticity makes increasingly notable contribution into material  behavior. The deviation of stress-strain relation in elastic regime from linear dependence is described by third-order (nonlinear) elastic moduli. In Murnaghan's theory \cite{Murnaghan} the nonlinear elastic behavior of an isotropic solid material is described by three nonlinear, third-order moduli ($l,m,n$) and two linear, second-order, Lam\'e constants ($\lambda, \mu$). It was demonstrated that third-order elastic moduli and their linear combinations are informative for the prediction of fatigue damage, for description of thermoelastic properties of crystalline solids, acoustic radiation stress, radiation-induced static strains, creep, thermal aging, wave processes, etc. In general nonlinear parameters were demonstrated to be more sensitive to structural changes in the material than linear ones.


In nanocomposites, the elastic nonlinear effects are enhanced due to localization of deformation near nanoparticles. 
Theoretical models of the nonlinear elastic properties of composite materials are still in an active development \cite{sevostianov2001,tsvelodub2000,tsvelodub2004,giordano2009, ColomboGiordano, giordano2017}. A theory, which takes into account nonlinear elastic properties of both the matrix and the filler was developed recently for the case of spherical inclusions \cite{Semenov_Beltukov2020}. The experimental validation of theoretical predictions is highly desirable, however by now measurements of nonlinear elastic properties of composites are very rare. And to the best of our knowledge no data on that for polymeric composites has been published as yet.

A wide variety of existing nanofillers can be classified basing on their dimensionality: 2D (nanosheets), 1D (nanotubes),  and 0D (spherical nanoparticles) \cite{bhattacharya2016}. As known carbon-based nanofillers exhibit most promising properties as potential nanofillers due to their high mechanical strength and high aspect ratio. For instance, carbon nanotubes were shown to have the tensile modulus up to 1 TPa, the tensile strength in the range of 50--150 GPa \cite{Wong1997,Yu2000} and high aspect ratio ($>1000$). Among 2D nanofillers Montmorillonite sheets provide relatively high Young's modulus reported to be within the range of 178-265 GPa \cite{Chen2006} and relatively high aspect ratio ($>50$). The excellent properties of these nanofillers make them good candidates for reinforcement of polymer matrices. However, the major drawbacks of high-aspect-ratio particles is an elevated agglomeration due to an increased surface-to-bulk ratio causing forces that attract particles to each other. Zero-dimensional particles demonstrate much lower agglomeration but provide moderate enhancement of mechanical properties. 

In this paper we report a complex multi-method experimental analysis of mechanical properties of composites based on polystyrene matrix with addition of nanofillers of different nature, dimensionality and size. 
Composite samples with different filler concentrations were fabricated by melt technology, and properties of both composite melts and resulting composite samples were examined and analyzed. 

The paper is organized as follows. Section 2 is devoted to the description of the applied materials, composite fabrication technology and methodologies used for testing of composite samples.  The results obtained and discussion are presented in Section 3. First we describe properties of composite melts obtained from thermal analysis by differential scanning calorimetry and control of filler distribution in the polymer matrix. Then we present data on linear elastic properties (elastic modulus, strength and strain at break) of composites obtained from tensile tests. Then we focus on those composites which demonstrated most profound changes of linear parameters as compared to pure polymer and present data on their nonlinear elastic parameters obtained from ultrasonic measurements at static stress. Finally we analyze changes in composites elasticity on the specific example of nonlinear elastic process: evolution of bulk strain solitary waves in fabricated materials.    
The conclusions made are summarized in Section 4.

\section{Materials and methods}
\label{S:2}

\subsection{Materials}

The grained 585 polystyrene (Nizhnekamskneftekhim, Russia)  was used as a polymer matrix for all the composite samples. The main parameters of polystyrene as specified by the manufacturer are shown in Table \ref{polystyrene}.  The following materials were applied as (nano)fillers: Silicon dioxide ($SiO_{2}$) particles Aerosil R812 modified by silazane (Evonic Industries, Germany); Alumina nanoparticles ($Al_{2}O_{3}$) Aeroxide Alu C805 modified by octylsilane (Evonic Industries, Germany); Carbon Black (CB) P-805E (Ivanovskiy tekuglerod and rubber, Russia), Multi-walled carbon nanotubes (CNT) CTube-100 (CNT Co. Ltd., Republic of Korea); Halloysite Natural Tubules (HNT) (NaturalNano Inc., USA), Sheet silicate (phyllosilicate) minerals Mica ME-100 (Mica) (CBC Co Ltd., Tokyo, Japan), Montmorillonite 15A (MMT) (Southern Clay Products Inc., USA). The data on filler particle dimensions are summarized in Table \ref{inclusions}.


As known \cite{Mokhireva2017,Devaraju2013} with an increase in the axial ratio of nanoparticles, an increase in the mechanical properties of polymer composites is observed at lower concentrations. So with the introduction of spherical dispersed particles, an increase in strength or elastic modulus of polymer composites is observed at filler concentrations more than tens of percent \cite{Giovino2018}, but, for example, with the introduction of carbon nanotubes, the increase of strength and stiffness of composites is already observed with the introduction a few percent or tenths of a percent \cite{Thostenson2002,Gojny2005}. It was shown in the works that in cases of intercalation of macromolecules and dispersion to individual nanolayers with a thickness of about 1 nm, the properties of polymer composites significantly improve already at low degrees of their filling with layered silicates \cite{Hari2017,Weng2016,Abenojar2017}. Therefore, in our work, spherical particles were introduced in PS-matrix up to 20\% with regard to polymer weight, layered silicates up to 5\% wt., anisodiametric filler concentration did not exceed 10-15\% wt.  

\begin{table}[h]
\centering
\begin{tabular}{l l}
\hline
\textbf{Parameter} & \textbf{Value} \\
\hline
Melt flow index & $2.8 \pm 0.7$  \\
Vicat softening temperature, ${}^{\circ}{\rm C}$ & 100  \\
Tensile strength at break, MPa & 48.0  \\
Flexural strength, MPa & 95.0 \\
Weight fraction of residual styrene, \% & 0.05   \\
\hline
\end{tabular}
\caption{Characteristics of 585 polystyrene}
\label{polystyrene}
\end{table}

\begin{table}[h]
\centering
\begin{tabular}{l l}
\hline
\textbf{Filler} & \textbf{Particle size} \\
\hline
$SiO_{2}$ & Diameter $\sim$ 7 nm  \\
$Al_{2}O_{3}$ & Diameter $\sim$ 13 nm  \\
CB & Diameter $\sim$ 80 nm  \\
CNT & Diameter 10--40 nm, \\
 & Length 1--25 {\textmu}m   \\
HNT & Diameter $\sim$ 100 nm,  \\
 & Length 0.5--1.2 {\textmu}m  \\
Mica & Average size 1--5 {\textmu}m  \\
MMT & Average size $\lesssim 10$ {\textmu}m \\
\hline
\end{tabular}
\caption{Dimensional specifications of filler particles}
\label{inclusions}
\end{table}

Nanocomposites consisting of the PS matrix with specified nanofillers were manufactured by melt technology. The affordability and efficient performance frequently make this technology a method of choice for polymer processing in various industries. 
PS-based compositions were prepared using a twin-screw micro compounder DSM Xplore 5 mL (Netherlands). Compounding was carried out at 220\,$^{\circ}{\rm C}$ for 10 min at 50 rpm/min. Block samples were fabricated by injecting the solution into a die heated to 80\,$^{\circ}{\rm C}$. When  removed from the microinjector the die self-cooled down to room temperature in air. Block samples of composites of two types have been manufactured: plates $50\times10\times1.5$ mm and blades with the working area of $20\times4\times1.5$ mm. Reference samples of the same shapes but made of pure PS have been manufactured as well for reference. 

\subsection{Testing of composite samples}

As known several key factors are critical for achieving an improvement in mechanical properties of a polymer through addition of a nanofiller. 
The following requirements should be addressed: (1) filler particles should have mechanical properties notably different from those of the matrix; (2) it is preferable that they would have high aspect ratio and high surface area to enable better interaction with the polymer; and (3) they should be well dispersed and agglomeration should be avoided \cite{bhattacharya2016}.

The comprehensive analysis was performed on mechanical properties of both PS-filler melts and composite samples. Rheological characteristics of melts of PS with different kinds of particles were determined using the rheometer Physica MCR 301 (Anton Paar GmbH, Austria) in a CP25-2 cone-plane measurement unit at 200\,$^{\circ}{\rm C}$ and 220\,$^{\circ}{\rm C}$ in shear and
dynamic (oscillatory) modes with a decrease (Down)
and increase (Top) of the strain rate (circular frequency) in air. 
The influence of particle concentration on the mechanical properties of nanocomposites in tension was studied using Instron 5940 universal testing system (USA) at a stretching rate of 5 mm/min and base length of 20 mm. Basing on the tensile test data, the strength at break ($\sigma{_b}$, MPa), strain at break ($\epsilon{_b}$, \%) and elastic modulus ($E{_0}$, GPa) have been determined.
The dispersion of nanoparticles in the polymer matrix was estimated from the micrographs of cryo-cleavage surfaces of the composite samples. The micrographs were taken using a Carl Zeiss Supra-55 scanning electron microscope.

\subsection{Evaluation of nonlinear elastic properties}

Nonlinear elastic properties of composite samples were examined  through relative variations of the third-order elastic moduli in nanocomosite samples as compared to those of pure polymer samples. The experimental methodology is based on the approach suggested by Hughes and Kelly  \cite{Hughes-Kelly1953} and utilizes the analysis of the dependence of the velocity of longitudinal and shear ultrasonic waves in the sample upon the applied static transverse stress. See \cite{PolymTest2020} for details of methodology applied.

Specimens were composed from 3 plates of a composite adhesively bonded with the Superglue ethylcyanoacrylate adhesive. Uniaxial static stress was applied to a longer side of the rectangular specimen. Testing was performed in perpendicular direction by longitudinal and shear sine ultrasonic waves at 2.25 MHz. Wave shifts at stepwise increase of the applied static stress were recorded
providing data on changes of wave velocity as function of applied stress.
Then the set of introduced effective longitudinal and shear moduli: $M_x = V_x^2 \rho_0$, $G_y = V_y^2 \rho_0$, $G_z = V_z^2 \rho_0$ plotted as function of stress provide data on the set of second- and third-order elastic moduli of the specimen material. The second-order Lame's moduli $\lambda,\mu$ are calculated from wave velocities at zero stress and third-order Murnaghan's moduli $l,m,n$ from the slope coefficients, as shown in \cite{PolymTest2020}.



The resulting equations for calculation of the set of Murnaghan's moduli can be written as:
\begin{gather}
    l = -\frac{3 \lambda +2 \mu }{2} \alpha _x-\frac{\lambda  (\lambda +\mu ) }{\mu}(1+2 \alpha
   _y)+\frac{\lambda ^2 }{2 \mu }(1-2 \alpha_z),\\
    m = -2 (\lambda +\mu ) \left(1+\alpha_y\right)+\lambda  \left(1-\alpha_z\right),\\
    n = -4 \mu  \left(1+\alpha _y-\alpha _z\right).
\end{gather}
where $\alpha_x$, $\alpha_y$ and $\alpha_z$ are dimensionless slope coefficients of the dependencies of corresponding effective moduli as function of applied uniaxial stress.

As known the nonlinear elastic moduli are more sensitive to structural changes in a material than the linear ones. Obviously, the third-order moduli of composites depend drastically not only on the type and concentration of inclusions but also on their distribution in the matrix. Inhomogeneous distribution and formation of agglomerates can cause considerable changes in these parameters. That means that ideally measurements of these parameters are to be made for each particular composite sample.

\subsection{Generation and monitoring of nonlinear strain waves}

The nonlinear elastic properties of materials
can provide conditions favourable for formation of
bulk strain solitary waves in waveguides made of
them.
As known (\cite{ams}), the nonlinear elasticity of a material can cause formation of a longitudinal strain wave $u$ with the amplitude $A$, velocity $v$ and width $L$, that is described by the equation:

\begin{equation}
u(x-vt) = A\cosh^{-2}\left(\frac{x-vt}{L}\right)  
\end{equation}

It can be shown that the relationship of the soliton amplitude with its velocity depends only on integral parameters of the material and takes the form:

\begin{equation}\label{soliton_ampl}
A = 3(v^2 - c^2) \frac{\int{\rho dy}}{\int{\beta dy}} 
\end{equation}
where $c$ is sound velocity and $\beta<0$ is the nonlinearity coefficient comprising a combination of the second- ($E,\nu$) and third-order Murnaghan ($l, m, n$) elastic moduli of the material:

\begin{equation}\label{beta}
\beta = 3E + 2l(1-2\nu)^3 +4m(1+\nu)^2(1-2\nu) +6n\nu^2  
\end{equation}

In our experiments solitons were formed in bar-shaped waveguides from initial laser-induced shock waves generated in water nearby the waveguide input. Soliton evolution in a waveguide was monitored using a digital holographic set-up as described in detail in our earlier papers, e.g. \cite{WaMot2017,spie2018}. 
The holographic approach provides information on spatial distributions of
refractive index gradient inside a transparent
sample and was already demonstrated to be an
efficient tool for detection of local density
variations formed by a strain wave (see \cite{strain2010,WaMot2017} and
references therein). 
Soliton amplitude and width were determined from
phase shift distributions introduced in the recording
wave front by wave-induced density gradients and its velocity was calculated from precise measurements of its position in the course of its propagation in the waveguide. 
Note that the reported holographic arrangement
operates with transparent specimens only, and to perform wave monitoring in opaque materials as all of the composites are, we recently suggested an amplificatory approach allowing
an indirect detection of strain waves in opaque
materials by monitoring phase shift gradients in a
layer of transparent material adhesively bonded to
the layer made of the material of interest
\cite{tpl2014,apl2018}. As shown in \cite{tpl2014}, in a layered bar made of two different materials a single soliton is formed which amplitude and width depend upon elastic properties of the corresponding materials. 
And as shown in \cite{apl2018}
the soliton velocity measured in such a sandwich
waveguide in one of the layers equals to the
arithmetic mean of soliton velocities in waveguides
of the same geometry but made of each of the
materials. 

Due to the integral dependence of soliton amplitude and velocity on the waveguide parameters (Eq.~(\ref{soliton_ampl})) these characteristics do not depend on the number and order of longitudinal layers of different materials in a waveguide. That is why in experiments we used two- and three-layered bars $10\times10$ mm in cross-section, made of a transparent layer of commercial PS and a layer(s) of a fabricated PS-based nanocomposite. Measurements made with these waveguides were compared with those made with a similar waveguide where the nanocomposite layer was substituted by a layer of pure PS but fabricated by the same technology as a composite. The nanocomposite layers and those of fabricated PS samples were made by bonding the corresponding plates with the ethylcyanoacrylate adhesive (Superglue). As we have demonstrated previously \cite{JAP2008,apl2018}, elastic features of this adhesive are close to those of the applied polymers and it does not affect noticeably the soliton behavior.


In these experiments we tested nanocomposite samples with the fillers of three types: spherical particles $SiO_2$, alumosilicate particles with high aspect ratio HNT, and carbon particles CB. Evolution of solitons in the waveguides with these materials was monitored and their velocities and decay decrements were determined.

\section{Results and discussion}
\label{S:3}

\subsection{Properties of composite melts}

\begin{figure}[t]
\centering
\includegraphics[width=10cm]{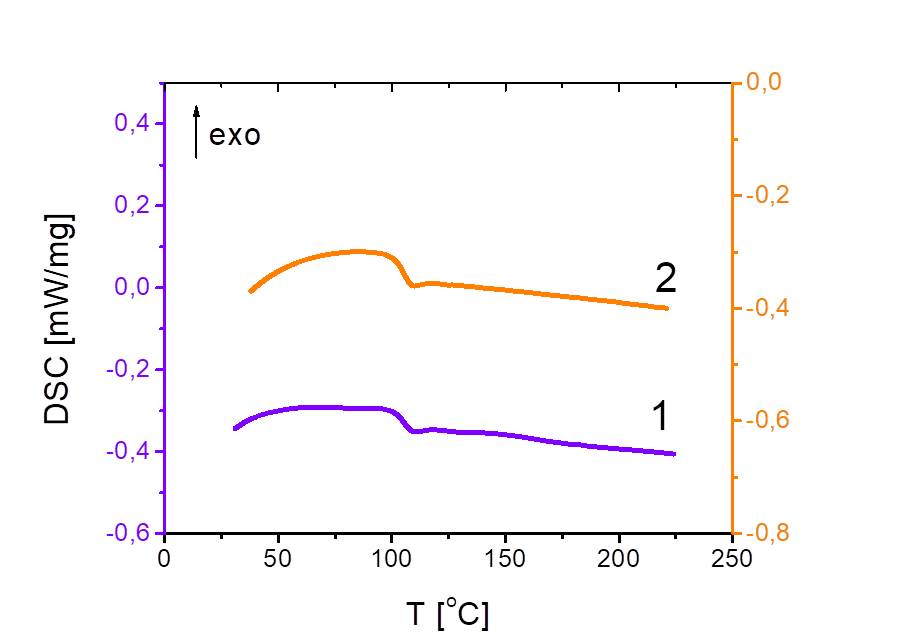}
\caption{DSC curve of pure PS pellets (1st and 2nd scan).}
\label{thermograms}
\end{figure}

Temperature ranges optimal for polystyrene
processing by melt technology, i.e. temperature in the extruder chamber and mold temperature, were evaluated by thermal analysis by differential scanning calorimetry (DSC). The glass transition temperature of pure PS pellets obtained from DSC thermograms (Fig.~\ref{thermograms}) comprised $T_g=106\,
^{\circ}{\rm C}$. Therefore the temperature range for polymer processing was chosen to be 200--220\,$
^{\circ}{\rm C}$, while the mold temperature had to be below 100\,$^{\circ}{\rm C}$.

\begin{figure}[t!]
\centering
\includegraphics[width=0.9\columnwidth]{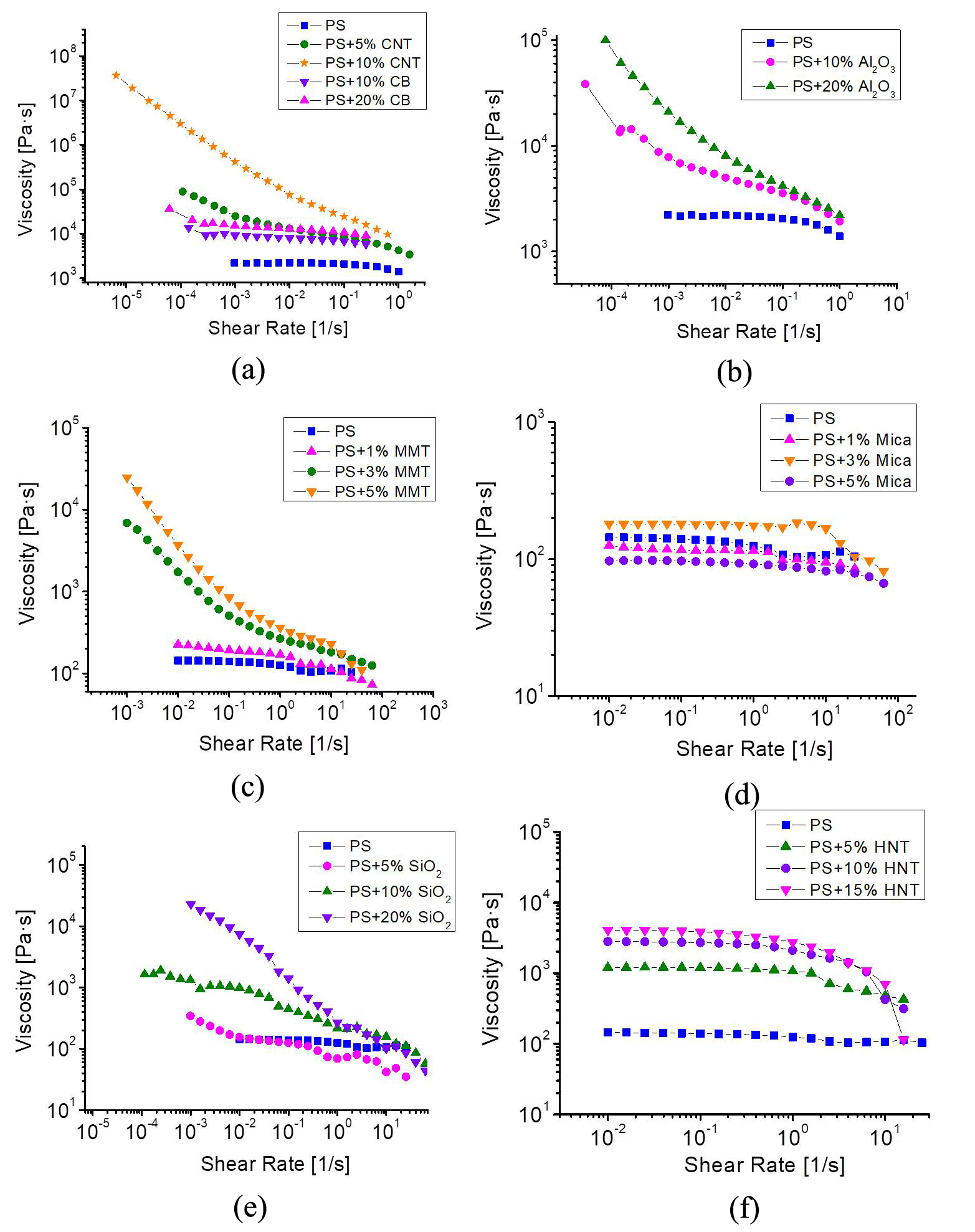}
\caption{Experimental dependencies of viscosity on
the deformation rate (circular frequency) for pure PS
and PS-based composites filled with carbon
nanoparticles at 200\,$^{\circ}{\rm C}$ (a), alumina
nanoparticles at 200\,$^{\circ}{\rm C}$ (b),
montmorillonite at 220\,$^{\circ}{\rm C}$ (c), sheet
silicate minerals at 220\,$^{\circ}{\rm C}$ (d),
silicon dioxide at 220\,$^{\circ}{\rm C}$ (e) and
halloysite natural tubules at 220\,$^{\circ}{\rm C}$
(f).}
\label{viscosity}
\end{figure}

Fig.~\ref{viscosity} presents experimental dependencies of melt viscosity on the deformation rate (circular frequency) for the fabricated PS-based composites with different filler concentrations.
As can be seen in Fig.~\ref{viscosity}(a) the
introduction of carbon nanoparticles caused an increase in melt viscosity, with the
highest rise observed for filler particles with high aspect ratio (CNT). The viscosity of composites with 10\% CB was higher than that of pure PS by an order of magnitude, while that of composites with 10\% CNT -- by two orders of magnitude at low shear rates. PS-based composites filled with
$Al_{2}O_{3}$  and MMT (Fig.~\ref{viscosity}(b,c)) showed similar dependencies: at  maximal filler concentrations, 20\% $Al_{2}O_{3}$ and 5\% MMT, the melt viscosity at low shear rates increased by two orders
of magnitude as compared to that of pure polymer. Composites filled with
$SiO_{2}$ and HNT also demonstrated increased 
melt viscosity, with the highest rise by an order of magnitude observed at low shear rates for samples with maximal filler concentrations, 20\% $SiO_{2}$ and 15\% HNT, see Fig.~\ref{viscosity}(e,f). The melt viscosity of composites with Mica particles remained about the same as that of pure PS, Fig.~\ref{viscosity}(d). 

Therefore, the introduction of different types of particles, except Mica, to PS melts caused noticeable rise of melt viscosity depending upon filler concentration. The most pronounced rise was observed with the introduction of MMT, CNT and $Al_{2}O_{3}$ at high concentrations. The effective viscosity of the polymer melt at these filler concentrations increased by an average of 2 or
5 orders of magnitude over the pure PS. The rise of melt viscosity at low shear rates
is known to be indicative of formation of a network structure between the polymer and a filler, see \cite{naira,gelfer}. 

\subsection{Filler distribution in the polymer matrix}

\begin{figure}[!t]
\centering
\includegraphics[width=0.9\columnwidth]{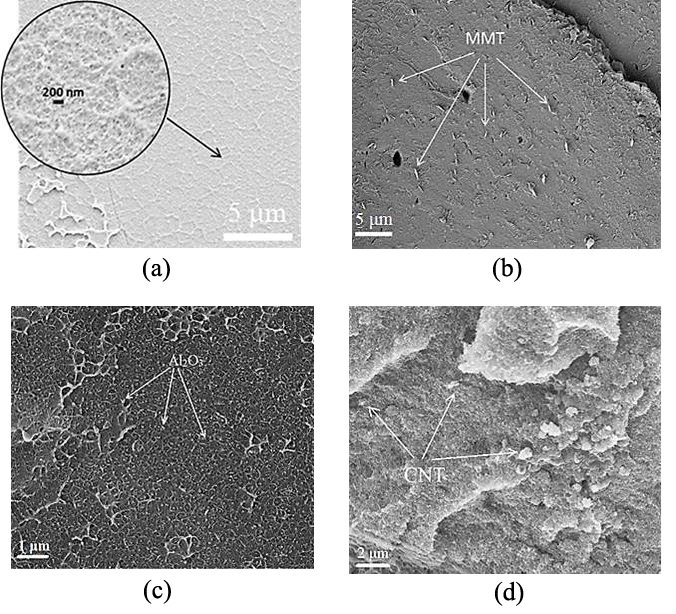}
\caption{Microphotographs of cryo-cleaved  surfaces of samples: a) pure PS, b) PS + 3\%
MMT, c) PS + 10\% $Al_2O_3$, d) PS + 10\% CNT.}
\label{microphoto}
\end{figure}

The dispersion of filler particles in the polymer matrix was controlled by microscopic analysis of cryo-cleavage surfaces of composites. Representative examples of microphotographs of composites containing 3\% MMT, 10\% $Al_2O_3$ and 10\% CNT are shown in Fig.~\ref{microphoto}. As can be seen in Fig.~\ref{microphoto} all the three types of particles were rather uniformly
distributed in the PS matrix, while agglomerates with a maximal size of about 
1--2 {\textmu}m were observed in some samples.

\subsection{Linear elastic properties of composites}

\begin{figure}[t]
\centering
\includegraphics[width=10cm]{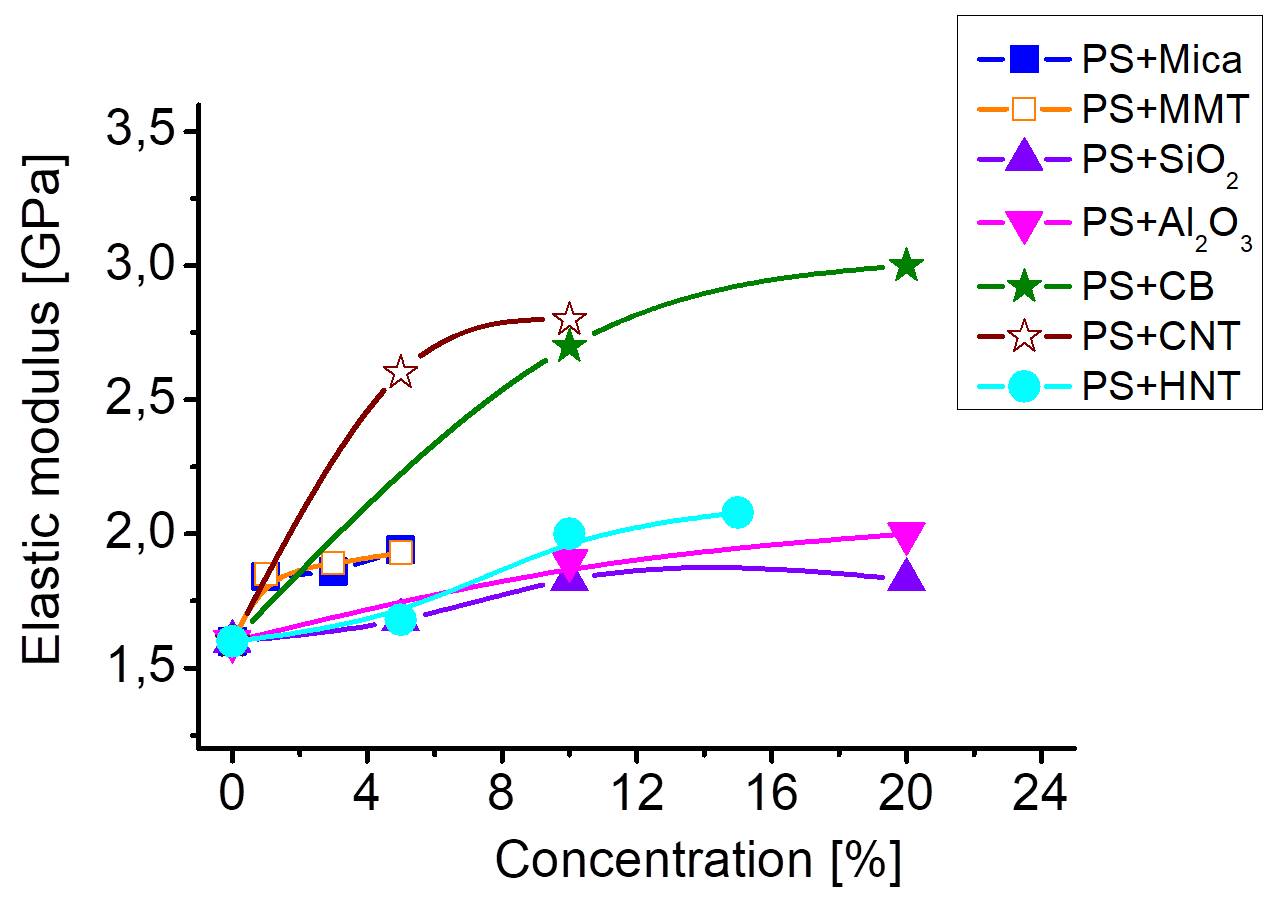}
\caption{Elastic modulus of PS-based composites as function of filler concentration.}
\label{moduli}
\end{figure}

The values of elastic modulus, strength and strain at break of fabricated samples of PS-based composites filled
with different nanoparticles are summarized in Table \ref{comp-parameters}. The dependence of the elastic modulus of 
composites on the type and concentration of fillers is presented in Fig.~\ref{moduli}.

As can be seen from Table \ref{comp-parameters}, the introduction of Mica nanoparticles led to an increase in the elastic
modulus of the PS-based composite reaching E = 1.9 GPa at the maximal concentration
of 5\%. That amounts for 19\% rise over that for pure PS ($E_0$ = 1.6 GPa). At the same time the elongation at break reduced by 30\% and reached a value of 3.9\%. It is worth noting
that composite strength remained at the level of pure PS ($\sigma_b$ = 56 MPa) regardless of Mica concentration.
The introduction of HNT also caused an
increase of the elastic modulus, however at higher concentrations of the filler. The noticeable rise of the modulus, in 25\% and 30\%, was achieved at 10\% and 15\% concentrations of HNT respectively. 
The elongation at break decreased by 60\%
and the strength at break by almost 20\% as compared to pure PS.
MMT nanoparticles provided changes in the elastic modulus similar to those with Mica at the same concentrations. While strength and elongation at break demonstrated more profound changes: at the concentration of 5\% MMT the sample strength
decreased from 56 to 46 MPa (18\%), and
elongation decreased from 5.6 to 2.6\% (54\%). Thus the introduction of different alumosilicate particles into the PS matrix can provided noticeable increase in elastic modulus with just small decrease of strength at break.   

Silicon oxide particles provided only slight increase in material rigidity at the concentrations up to 10\%,
while concentrations over 10\% did not cause any further significant rise of the elastic modulus that achieved a maximum value of 1.83 GPa (14\% rise over pure PS). With increasing
stiffness samples became much more brittle, their deformation at break $\epsilon{_b}$ dropped down by 60\% and achieved the value of 2.2\% at 20\% concentration of $SiO_{2}$. The sample strength also decreased by about 50\% at the concentrations of 10\% and 20\% $SiO_{2}$. Introduction of spherical nanoparticles of
$Al_{2}O_{3}$  resulted in the increase of material stiffness in tension. The elastic modulus rised up to 2 GPa (20\% rise) at the filler concentration of 20\%.  At the same
time, the composite strength and strain at beak decreased by 20\% and 40\%, respectively. Thus, the introduction of small spherical particles did not provide any challenging improvement of elastic properties of material.

The introduction of carbon nanoparticles caused most pronounced changes in elastic properties of composites. In particular, stiffness in tension of PS-based composites
filled with CB increased noticeably and at the concentration
of 10\% the elastic modulus reached $E_0=2.8$ GPa, that is 65\% higher than that for pure PS. However higher filler concentrations did not cause any significant increase of the elastic modulus, which   rised up to $E_0=3.0$ GPa only at 20\% CB. With increasing stiffness of
the samples deformation at break decreased and at the concentration of 20\%
$\epsilon{_b}$ comprised 2.1\%, which is 2.5 times lower than that of pure PS. The strength at break reached its maximum of 74 MPa at 10\% of CB, rising by 17\% over that of pure PS and then dropped down to 65 MPa at 20\% CB, being however still higher than that of pure PS.
The introduction of CNT also provided an increase in the elastic modulus of composites, which achieved 2.8 GPa at 10\% concentration. 
At the same time particles with high aspect ratio caused the decrease in both composite strength and strain at break. At 10\% concentration of CNT $\sigma{_b}$ = 49 MPa, that is 12\% lower than that of pure PS and $\epsilon{_b}$ = 1.7\%, that is more than 3 times lower than that of pure PS.

\begin{table}[h]
\centering
\begin{tabular}{l l l l}
\hline
\textbf{Sample} & \textbf{Strength} & \textbf{Tensile elastic} & \textbf{Strain}\\
                & \textbf{at break} & \textbf{modulus}         & \textbf{at break}\\
                & $\sigma{_b}$, MPa & $E$, GPa                 & $\epsilon_b$, \%\\
\hline
PS pure               & $56\pm1$ & $1.6 \pm0.1 $ & $5.6\pm0.2$\\
PS+1\% Mica           & $56\pm4$ & $1.84\pm0.02$ & $4.4\pm0.1$\\
PS+3\% Mica           & $54\pm2$ & $1.86\pm0.04$ & $4.1\pm0.2$\\
PS+5\% Mica           & $56\pm5$ & $1.94\pm0.05$ & $3.9\pm0.3$\\
PS+5\% HNT            & $52\pm1$ & $1.68\pm0.15$ & $5.7\pm0.3$\\
PS+10\% HNT           & $49\pm6$ & $2.00\pm0.06$ & $2.8\pm0.5$\\
PS+15\% HNT           & $46\pm4$ & $2.08\pm0.09$ & $2.4\pm0.2$\\
PS+1\% MMT            & $58\pm4$ & $1.85\pm0.04$ & $3.9\pm0.4$\\
PS+3\% MMT            & $50\pm4$ & $1.89\pm0.04$ & $2.9\pm0.3$\\
PS+5\% MMT            & $46\pm2$ & $1.93\pm0.07$ & $2.6\pm0.2$\\
PS+5\% $SiO_{2}$      & $52\pm2$ & $1.68\pm0.15$ & $5.7\pm0.3$\\
PS+10\% $SiO_{2}$     & $32\pm1$ & $1.83\pm0.17$ & $2.8\pm0.2$\\
PS+20\% $SiO_{2}$     & $31\pm3$ & $1.83\pm0.2 $ & $2.2\pm0.2$\\
PS+10\% $Al_{2}O_{3}$ & $57\pm1$ & $1.9 \pm0.1 $ & $4.1\pm0.5$\\
PS+20\% $Al_{2}O_{3}$ & $49\pm1$ & $2.0 \pm0.1 $ & $3.1\pm0.2$\\
PS+10\% CB            & $74\pm2$ & $2.8 \pm0.1 $ & $3.1\pm0.2$\\
PS+20\% CB            & $65\pm5$ & $3.0 \pm0.2 $ & $2.1\pm0.6$\\
PS+5\% CNT            & $53\pm3$ & $2.6 \pm0.2 $ & $1.9\pm0.3$\\
PS+10\% CNT           & $49\pm3$ & $2.8 \pm0.2 $ & $1.7\pm0.2$\\
\hline
\end{tabular}
\caption{Mechanical properties of PS-based composites with different fillings.}
\label{comp-parameters}
\end{table}

The behavior of elastic modulus of composites plotted in Fig.~\ref{moduli} as function of filler concentration demonstrates that introduction of more rigid particles leads to more profound increase in the elastic modulus, with the highest rise obtained with carbon nanofillers. The elastic modulus rised from 1.6 GPa for pure PS to 2.8 GPa for composites with 10\% of CNT or CB. It should be emphasized that CB particles provided higher strength and strain at break of composites than CNT, with strength at break being even higher than that of pure PS. We can thus conclude that from the point of view of linear elasticity carbon nanofillers provide more pronounced changes in elastic properties, with CB seeming more promising.

\subsection{Nonlinear elastic properties}

The analysis of nonlinear elastic properties of fabricated composite samples was performed for samples from each group of nanofillers, which demonstrated most profound changes of linear elastic properties: PS + 20\% SiO${}_2$, PS + 10\% HNT and PS + 20\% CB. The analysis was based on measurements of the third-order elastic moduli and examined in experiments on monitoring of the evolution of bulk nonlinear strain waves in these composites.

\subsubsection{Measurements of the third-order elastic moduli}

Table \ref{nonlinear moduli} presents sets of data on the second- and third-order moduli for 3-plate layered sandwiches of these composites, obtained from ultrasonic measurements. For comparison measurements were also performed for a bulk non-layered specimen made of a commercial grade pure polystyrene.  

As can be seen from the Table elastic parameters of the specimen made of adhesively bonded plates of pure polystyrene fabricated by melt technology differ from those of commercial grade bulk specimen only slightly, and can be due to several factors, among which are the difference in polymer structure and fabrication process and the influence of adhesive layers.  
At the same time as can be readily seen from the Table all the three nanofillers cause more noticeable changes in both linear and nonlinear elastic moduli. The noticeable rise of the second-order Lame moduli $\lambda$ and $\mu$ is observed in all composites
The Young's modulus $E$ calculated from these data comprised 3.85 GPa for PS pure and
4.23, 4.29 and 4.58 GPa 
for nanocomposites with SiO${}_2$, HNT and CB particles respectively. Thus in terms of linear elastic properties the addition of studied nanofillers to the polystyrene matrix provided efficient reinforcement of the material. The Young's modulus $E$ of composites obtained by ultrasonic measurements exceeded the static one (Table \ref{comp-parameters}) by approximately 2 GPa. It corresponds to a typical frequency dependence of elastic moduli of polymer materials \cite{Yadav-2020}.

Changes in the nonlinear, third-order moduli $l$, $m$ and $n$, demonstrate more complex behavior. In general for all the composite samples  changes in the nonlinear moduli were more profound than changes in the linear, second-order moduli $\lambda$ and $\mu$.
Among the nonlinear moduli the $l$ modulus demonstrated  prominent variations in all the composites with the maximal change of about 84 \% observed in the CB-containing one. The $n$ modulus showed high relative variation, of about 91 \% only in PS+HNT samples, while variations of the $m$ modulus were not so high and reached about 50\% 
in PS+HNT and PS+CB nanocomposites.

\begin{table}[h]
\centering
\begin{tabular}{lccccc}
\hline
\textbf{Material}  & \multicolumn{5}{c}{\textbf{Elastic moduli, GPa}} \\
&  $\lambda$ & $\mu$ & $l$  & $m$  &  $n$
\\
\hline
PS commercial      & 2.80  &  1.44 & $-46.2 \pm 1.8$ & $-14.8 \pm 0.7$ &  $-7.5 \pm 0.6$ \\
PS pure            & 2.76  & 1.45  & $-44.5 \pm 1.3$ & $-12.8 \pm 0.5$ & $-5.7  \pm 0.4$  \\
PS + 20\% $SiO_2$ & 3.35  & 1.58  &  $-64.2 \pm 1.9$ & $-15.8 \pm 0.9$ & $-7.35 \pm 0.7$ \\
PS + 10\% HNT     & 3.02  & 1.62  &  $-40.9 \pm 2.1$ & $-18.6 \pm 1.0$ & $-10.9 \pm 0.9$ \\
PS + 20\% CB      & 3.58  & 1.71  &  $-81.9 \pm 2.7$ & $-18.9 \pm 0.8$ & $-5.4  \pm 0.8$ 
\\
\hline
\end{tabular}
\caption{Second- and third-order elastic moduli of nanocomposite samples determined from ultrasonic measurements. Second-order moduli $\lambda$ and $\mu$ is measured within error tolerance of 0.02 and 0.01 GPa, respectively.}
\label{nonlinear moduli}
\end{table}

\subsubsection{Evolution of bulk strain solitons}

The effect of changes in elastic features of polystyrene provided by addition of different nanofillers was examined on the particular example of nonlinear elastic process: formation and evolution of bulk nonlinear strain waves in two- or three-layered waveguides containing one or two nanocomposite layer(s), respectively. In common with ultrasonic measurements experiments were performed with the three types of fabricated composites: PS + 20\% $SiO_2$ particles, PS + 10\% HNT and PS + 20\% CB, which demonstrated most profound change of the elastic modulus among each type of fillers (see Table \ref{comp-parameters}). Soliton parameters obtained in waveguides with composite layers were compared with those obtained in waveguides of the same construction but with layers of pure polystyrene instead of composites.



\begin{figure}[t]
\centering
\includegraphics[width=\linewidth]{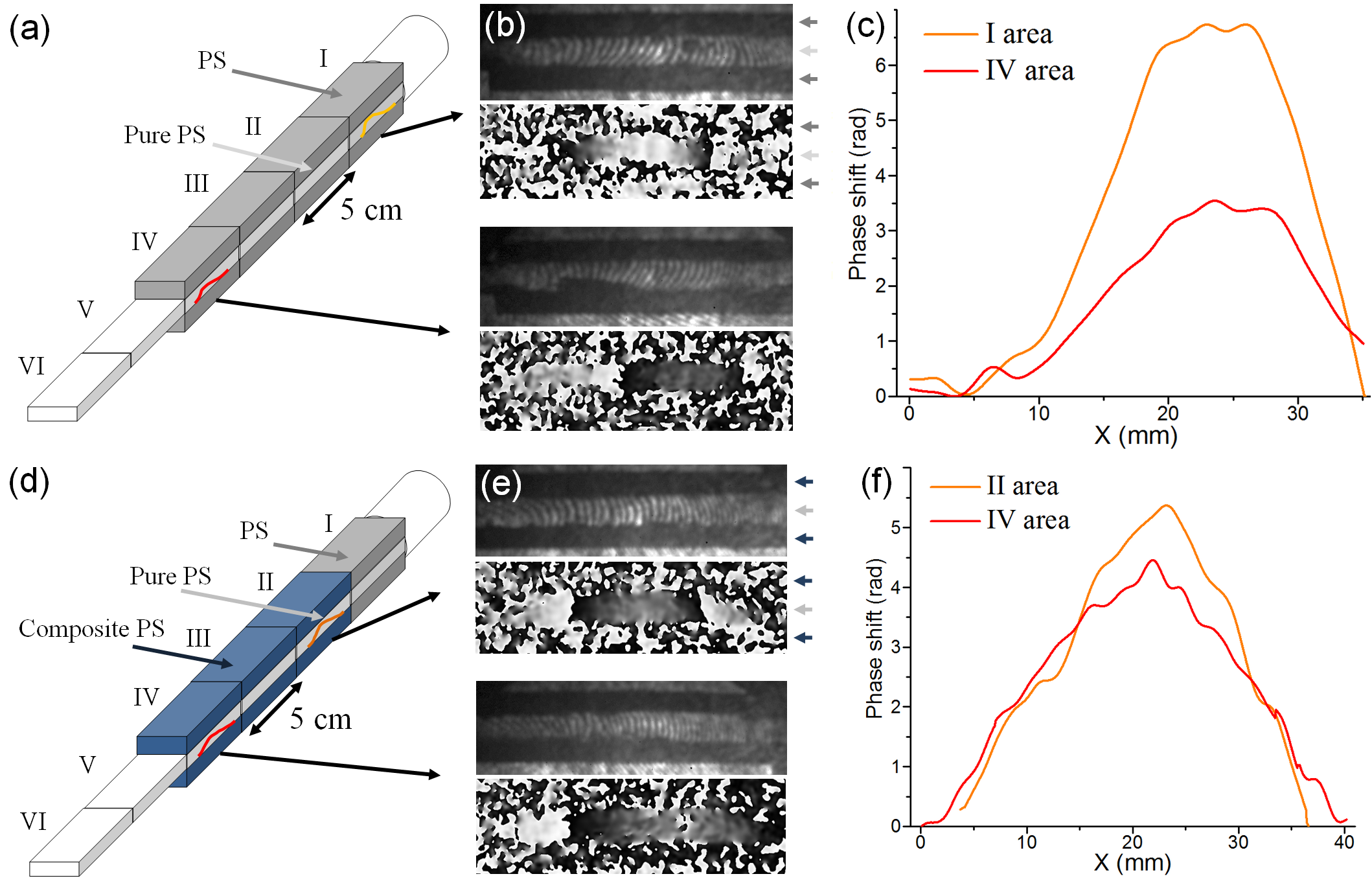}
\caption{Holographic recording of solitons propagating in 3-layered waveguides made of pure PS (a--c) and  with layers of nanocomposite PS+10\%HNT (d--f). (a,d) -- waveguide schematics; (b, e) -- digital holograms and reconstructed phase images of solitons in the beginning (top rows) and at the end (bottom rows) of layered structures; 
(c,f) - phase shift distributions in corresponding solitons.}
\label{soliton}
\end{figure}

Fig.~\ref{soliton} presents waveguide schematics for the 3-layer layout, representative examples of recorded digital holograms and reconstructed phase images obtained in waveguides with HNT-containing composite layers and phase shift distributions representing solitons in the transparent middle layer.

Since the soliton is essentially a long trough-shaped wave with smooth fronts, its width is measured with relatively high error. Therefore our analysis was based on the data on soliton velocity (which is directly related to its amplitude) and decay decrement. Taking into account the
direct proportion between the strain wave amplitude and recorded phase shift, the decay decrement was
calculated as: 

\begin{equation}\label{Alpha-measurement}
\alpha = \frac{1}{x_2 - x_1}\ln\frac{\varphi_1}{\varphi_2}
\end{equation}
where $x_1$ and $x_2$ are positions of soliton maxima in the two areas of the waveguide (for example, in zones I and IV in Fig.~\ref{soliton}(a) and in zones II and IV in Fig.~\ref{soliton}(d)).

The data obtained on soliton parameters in the above mentioned waveguides is summarized in Table \ref{soliton data}. As can be seen from the Table, silica nanoparticles caused small modifications to the soliton parameters that are almost within the experimental errors. HNTs provided more noticeable changes, especially in terms of the decreased decay decrement. And addition of CB particles resulted in most profound changes in soliton parameters. The general trend of these changes correlates with changes of elastic  modulus obtained for composites with these fillings from tensile tests (Table \ref{comp-parameters}). 
Furthermore, the nonlinearity coefficient $\beta$ (Eq.~(\ref{beta})), calculated from the data on linear and nonlinear elastic moduli obtained from ultrasonic measurements (see Table \ref{soliton data}),   demonstrated similar trend, showing least change in $SiO_2$-containing composites, and more significant changes in composites with HNT and CB particles.   

\begin{table}[h]
\centering
\begin{tabular}{l l l l}
\hline
\textbf{Material} & \textbf{Non-} &  \textbf{Soliton} & \textbf{Decay} \\
 & \textbf{linearity} &  \textbf{velocity} & \textbf{decrement} \\
& $\beta,$ GPa & $v$, m/s & $\alpha$, cm$^{-1}$\\
\hline
PS commercial     & $-32.6$ & $1800\pm7$ & $0.012 \pm0.006$ \\
PS pure           & $-26.8$ & $1772\pm10$ & $0.041 \pm0.006$ \\
PS + 20\% $SiO_2$ & $-33.0$ & $1801\pm9$ & $0.039 \pm0.006$  \\
PS + 10\% HNT     & $-43.2$ & $1820\pm10$ & $0.018 \pm0.004$\\
PS + 20\% CB      & $-39.3$ & $1887\pm13$ & $0.010 \pm0.005$\\
\hline
\end{tabular}
\caption{Soliton velocities and decay decrements in nanocomposites.}
\label{soliton data}
\end{table}


\section{Conclusions}
\label{S:4}

We have thus performed a comprehensive analysis of mechanical properties of PS-based nanocomposites filled with different types of inclusions: spherical particles ($SiO_{2}$ and $Al_{2}O_{3}$), alumosilicates (montmorillonite, halloysite natural tubules and Mica) and carbon fillers (carbon black and multiwalled carbon nanotubes). The analysis was performed on composite samples fabricated under the same conditions and 
comprised tests at all steps of material fabrication procedure followed by assessment of linear elastic properties using standard measurement technologies and of nonlinear elastic parameters using the laboratory set-up. The composite behavior was finally evaluated by monitoring of evolution of nonlinear waves in it. 

Testing of composite melts showed that introduction of all types of particles, except Mica, to PS melts caused noticeable rise of melt viscosity being more profound at higher filler concentration. The most pronounced rise was observed with the introduction of MMT, CNT and $Al_{2}O_{3}$ at high concentrations. At low shear rates the effective viscosity of polymer melts at these filler concentrations increased by an average of 2 to 5 orders of magnitude over pure PS, that is known to be indicative of a network structure formed by the polymer and the filler.

The control of filler distribution in the polymer matrix verified that particles of all types were sufficiently uniformly distributed in the polymer matrix, agglomerates sized no bigger than 1--2 {\textmu}m have been observed in some samples at higher filler concentrations. 

The analysis of linear elastic properties of fabricated samples demonstrated that introduction of more rigid particles led to more profound increase in the elastic modulus, with the highest rise of about 80\% obtained with carbon fillers. It is worth noting that along with the enhancement of elastic modulus CB particles provided also enhanced strength at break, about 20\% higher than that of pure PS, and CNT particles allowed for maintaining this value about the same as that of pure PS. 

Measurements of nonlinear elastic parameters demonstrated that the third-order elastic moduli are more sensitive to addition of filler particles to the polymer matrix than the second-order ones. However, no general trend was observed in their variations from one filler to another. Alternatively the nonlinearity coefficient $\beta$ used for description of bulk nonlinear strain waves and comprising the combination of linear and nonlinear elastic moduli of a material demonstrated considerable changes correlating with changes of the Young's modulus in these composites. The absolute value of $\beta$ showed highest rise in 1.6 times in the HNT-containing composite as compared to that in pure PS.  

The changes in nonlinear elasticity of fabricated composites were validated by measurements of the velocity and decay decrements of bulk nonlinear strain waves. It was shown that the higher was rise in ${\vert}\beta{\vert}$ value the more significant was the increase of wave velocity and decrease of its decay decrement.

We have to note that to the best of our knowledge this is the first so fully-featured research of the complete set of linear and nonlinear elastic properties performed for  polymer nanocomposites with different types of filler particles.

\section{Acknowledgments}
\label{S:5}

The financial support from the Russian Science Foundation
under the grant \# 17-72-20201 is gratefully
acknowledged.





\bibliographystyle{model1-num-names}
\bibliography{sample.bib}

\begin{thebibliography}{44}
\expandafter\ifx\csname natexlab\endcsname\relax\def\natexlab#1{#1}\fi
\providecommand{\bibinfo}[2]{#2}
\ifx\xfnm\relax \def\xfnm[#1]{\unskip,\space#1}\fi
\bibitem[{Ferreira et~al.(2016)Ferreira, Nóvoa, and Marques}]{multifunc2016}
\bibinfo{author}{A.~D. B.~L. Ferreira}, \bibinfo{author}{P.~R.~O. Nóvoa},
  \bibinfo{author}{A.~T. Marques},
\newblock \bibinfo{title}{Multifunctional material systems: A state-of-the-art
  review},
\newblock \bibinfo{journal}{Composite Structures} \bibinfo{volume}{151}
  (\bibinfo{year}{2016}) \bibinfo{pages}{3--35}.
\bibitem[{Gibson(2010)}]{gibson}
\bibinfo{author}{R.~F. Gibson},
\newblock \bibinfo{title}{A review of recent research on mechanics of
  multifunctional composite materials and structures},
\newblock \bibinfo{journal}{Composite Structures} \bibinfo{volume}{92}
  (\bibinfo{year}{2010}) \bibinfo{pages}{2793--2810}.
\bibitem[{Mittal(2010)}]{Optimiz_2010}
\bibinfo{editor}{V.~Mittal} (Ed.), \bibinfo{title}{Optimization of Polymer
  Nanocomposite Properties}, \bibinfo{publisher}{WILEY-VCH Verlag GmbH \& Co.},
  \bibinfo{year}{2010}.
\bibitem[{Koratkara et~al.(2005)Koratkara, Suhr, Joshi, Kane, Schadler, Ajayan,
  and Bartolucci}]{Koratkara2005}
\bibinfo{author}{N.~A. Koratkara}, \bibinfo{author}{J.~Suhr},
  \bibinfo{author}{A.~Joshi}, \bibinfo{author}{R.~S. Kane},
  \bibinfo{author}{L.~S. Schadler}, \bibinfo{author}{P.~M. Ajayan},
  \bibinfo{author}{S.~Bartolucci},
\newblock \bibinfo{title}{Characterizing energy dissipation in single-walled
  carbon nanotube polycarbonate composites},
\newblock \bibinfo{journal}{Applied Physics Letters} \bibinfo{volume}{87}
  (\bibinfo{year}{2005}) \bibinfo{pages}{063102}.
\bibitem[{Uddin and Sun(2008)}]{Uddin2008}
\bibinfo{author}{M.~F. Uddin}, \bibinfo{author}{C.~T. Sun},
\newblock \bibinfo{title}{Strength of unidirectional glass/epoxy composite with
  silica nanoparticle-enhanced matrix},
\newblock \bibinfo{journal}{Composites Science and Technology}
  \bibinfo{volume}{68} (\bibinfo{year}{2008}) \bibinfo{pages}{1637--1643}.
\bibitem[{Bhattacharya(2016)}]{bhattacharya2016}
\bibinfo{author}{M.~Bhattacharya},
\newblock \bibinfo{title}{Polymer nanocomposites - a comparison between carbon
  nanotubes, graphene, and clay as nanofillers},
\newblock \bibinfo{journal}{Materials} \bibinfo{volume}{9}
  (\bibinfo{year}{2016}) \bibinfo{pages}{262}.
\bibitem[{Papageorgiou et~al.(2017)Papageorgiou, Kinloch, and
  Young}]{Graphene2017}
\bibinfo{author}{D.~G. Papageorgiou}, \bibinfo{author}{I.~A. Kinloch},
  \bibinfo{author}{R.~J. Young},
\newblock \bibinfo{title}{Mechanical properties of graphene and graphene-based
  nanocomposites},
\newblock \bibinfo{journal}{Progress in Materials Science} \bibinfo{volume}{90}
  (\bibinfo{year}{2017}) \bibinfo{pages}{75--127}.
\bibitem[{Li et~al.(2019)Li, Huang, Zeng, Li, Tian, Fu, Wang, and
  Zhong}]{review-carbon2019}
\bibinfo{author}{Y.~Li}, \bibinfo{author}{X.~Huang}, \bibinfo{author}{L.~Zeng},
  \bibinfo{author}{R.~Li}, \bibinfo{author}{H.~Tian}, \bibinfo{author}{X.~Fu},
  \bibinfo{author}{Y.~Wang}, \bibinfo{author}{W.-H. Zhong},
\newblock \bibinfo{title}{A review of the electrical and mechanical properties
  of carbon nanofiller-reinforced polymer composites},
\newblock \bibinfo{journal}{Journal of Materials Science} \bibinfo{volume}{54}
  (\bibinfo{year}{2019}) \bibinfo{pages}{1036--1076}.
\bibitem[{Eshelby(1957)}]{eshelby1957}
\bibinfo{author}{J.~Eshelby},
\newblock \bibinfo{title}{The determination of the elastic field of an
  ellipsoidal inclusion, and related problems},
\newblock \bibinfo{journal}{Proceedings of the Royal Society of London. Series
  A. Mathematical and Physical Sciences} \bibinfo{volume}{241}
  (\bibinfo{year}{1957}) \bibinfo{pages}{376--396}.
\bibitem[{Halpin and Kardos(1976)}]{halpin-kardos76}
\bibinfo{author}{J.~Halpin}, \bibinfo{author}{J.~Kardos},
\newblock \bibinfo{title}{The \uppercase{H}alpin-\uppercase{T}sai equations: a
  review},
\newblock \bibinfo{journal}{Polymer Engineering and Science}
  \bibinfo{volume}{16} (\bibinfo{year}{1976}) \bibinfo{pages}{344--352}.
\bibitem[{Matveeva and van Hattum(2011)}]{matveeva11}
\bibinfo{author}{A.~Matveeva}, \bibinfo{author}{F.~van Hattum},
\newblock \bibinfo{title}{Design and analysis of structural models of composite
  materials based on carbon nanotubes},
\newblock \bibinfo{journal}{Proc. Intern. Conf. "Modern Problems of Applied
  Mathematics and Mechanics", May 30-June 4, 2011, Novosibirsk, Russia}
  (\bibinfo{year}{2011}).
\bibitem[{Mori and Tanaka(1973)}]{mori-tanaka}
\bibinfo{author}{T.~Mori}, \bibinfo{author}{K.~Tanaka},
\newblock \bibinfo{title}{Average stress in matrix and average elastic energy
  of materials with misfitting inclusions},
\newblock \bibinfo{journal}{Acta Metallurgica} \bibinfo{volume}{21}
  (\bibinfo{year}{1973}) \bibinfo{pages}{571--574}.
\bibitem[{Ji. et~al.(2010)Ji., Cao, and Feng}]{ji-cao-feng}
\bibinfo{author}{X.~Ji.}, \bibinfo{author}{Y.~Cao}, \bibinfo{author}{X.~Feng},
\newblock \bibinfo{title}{Micromechanics prediction of the effective elastic
  moduli of graphene sheet-reinforced polymer nanocomposites},
\newblock \bibinfo{journal}{Modelling and Simulation in Materials Science and
  Engineering} \bibinfo{volume}{18} (\bibinfo{year}{2010})
  \bibinfo{pages}{1--16}.
\bibitem[{Murnaghan(1951)}]{Murnaghan}
\bibinfo{author}{E.~D. Murnaghan}, \bibinfo{title}{Finite Deformation of an
  Elastic Solid}, \bibinfo{publisher}{Wiley, New York}, \bibinfo{year}{1951}.
\bibitem[{Sevostianov and Vakulenko(2001)}]{sevostianov2001}
\bibinfo{author}{I.~Sevostianov}, \bibinfo{author}{A.~Vakulenko},
\newblock \bibinfo{title}{Inclusion with nonlinear properties in elastic
  medium},
\newblock \bibinfo{journal}{International Journal of Fracture}
  \bibinfo{volume}{107} (\bibinfo{year}{2001}) \bibinfo{pages}{9--14}.
\bibitem[{Tsvelodub(2000)}]{tsvelodub2000}
\bibinfo{author}{I.~Y. Tsvelodub},
\newblock \bibinfo{title}{Determination of the strength characteristics of a
  physically nonlinear inclusion in a linearly elastic medium},
\newblock \bibinfo{journal}{Journal of Applied Mechanics and Technical Physics}
  \bibinfo{volume}{41} (\bibinfo{year}{2000}) \bibinfo{pages}{734--739}.
\bibitem[{Tsvelodub(2004)}]{tsvelodub2004}
\bibinfo{author}{I.~Y. Tsvelodub},
\newblock \bibinfo{title}{Physically nonlinear ellipsoidal inclusion in a
  linearly elastic medium},
\newblock \bibinfo{journal}{Journal of Applied Mechanics and Technical Physics}
  \bibinfo{volume}{45} (\bibinfo{year}{2004}) \bibinfo{pages}{69--75}.
\bibitem[{Giordano et~al.(2009)Giordano, Palla, and Colombo}]{giordano2009}
\bibinfo{author}{S.~Giordano}, \bibinfo{author}{P.~Palla},
  \bibinfo{author}{L.~Colombo},
\newblock \bibinfo{title}{Nonlinear elasticity of composite materials},
\newblock \bibinfo{journal}{The European Physical Journal B}
  \bibinfo{volume}{68} (\bibinfo{year}{2009}) \bibinfo{pages}{89}.
\bibitem[{Colombo and Giordano(2011)}]{ColomboGiordano}
\bibinfo{author}{L.~Colombo}, \bibinfo{author}{S.~Giordano},
\newblock \bibinfo{title}{Nonlinear elasticity in nanostructured materials},
\newblock \bibinfo{journal}{Reports on Progress in Physics}
  \bibinfo{volume}{74} (\bibinfo{year}{2011}) \bibinfo{pages}{116501}.
\bibitem[{Giordano(2017)}]{giordano2017}
\bibinfo{author}{S.~Giordano},
\newblock \bibinfo{title}{Nonlinear effective properties of heterogeneous
  materials with ellipsoidal microstructure},
\newblock \bibinfo{journal}{Mechanics of Materials} \bibinfo{volume}{105}
  (\bibinfo{year}{2017}) \bibinfo{pages}{16--28}.
\bibitem[{Semenov and Beltukov(2020)}]{Semenov_Beltukov2020}
\bibinfo{author}{A.~A. Semenov}, \bibinfo{author}{Y.~M. Beltukov},
\newblock \bibinfo{title}{Nonlinear elastic moduli of composite materials with
  nonlinear spherical inclusions dispersed in a nonlinear matrix},
\newblock \bibinfo{journal}{International Journal of Solids and Structures}
  (\bibinfo{year}{2020}).
\bibitem[{Wong et~al.(1997)Wong, Sheehan, and Lieber}]{Wong1997}
\bibinfo{author}{E.~W. Wong}, \bibinfo{author}{P.~E. Sheehan},
  \bibinfo{author}{C.~M. Lieber},
\newblock \bibinfo{title}{Nanobeam mechanics: Elasticity, strength, and
  toughness of nanorods and nanotubes},
\newblock \bibinfo{journal}{Science} \bibinfo{volume}{277}
  (\bibinfo{year}{1997}) \bibinfo{pages}{1971--1975}.
\bibitem[{Yu et~al.(1997)Yu, Lourie, Dyer, Moloni, Kelly, and Ruoff}]{Yu2000}
\bibinfo{author}{M.-F. Yu}, \bibinfo{author}{O.~Lourie}, \bibinfo{author}{M.~J.
  Dyer}, \bibinfo{author}{K.~Moloni}, \bibinfo{author}{T.~F. Kelly},
  \bibinfo{author}{R.~S. Ruoff},
\newblock \bibinfo{title}{Strength and breaking mechanism of multiwalled carbon
  nanotubes under tensile load},
\newblock \bibinfo{journal}{Science} \bibinfo{volume}{287}
  (\bibinfo{year}{1997}) \bibinfo{pages}{637--640}.
\bibitem[{Chen and Evans(2006)}]{Chen2006}
\bibinfo{author}{B.~Chen}, \bibinfo{author}{J.~R. Evans},
\newblock \bibinfo{title}{Elastic moduli of clay platelets},
\newblock \bibinfo{journal}{Scripta Materialia} \bibinfo{volume}{54}
  (\bibinfo{year}{2006}) \bibinfo{pages}{1581--1585}.
\bibitem[{Mokhireva et~al.(2017)Mokhireva, Svistkov, Solod'ko, Komar, and
  Stöckelhuber}]{Mokhireva2017}
\bibinfo{author}{K.~A. Mokhireva}, \bibinfo{author}{A.~L. Svistkov},
  \bibinfo{author}{V.~N. Solod'ko}, \bibinfo{author}{L.~A. Komar},
  \bibinfo{author}{K.~W. Stöckelhuber},
\newblock \bibinfo{title}{Experimental analysis of the effect of carbon
  nanoparticles with different geometry on the appearance of anisotropy of
  mechanical properties in elastomeric composites},
\newblock \bibinfo{journal}{Polymer Testing} \bibinfo{volume}{59}
  (\bibinfo{year}{2017}) \bibinfo{pages}{46--54}.
\bibitem[{Devaraju et~al.(2013)Devaraju, Kumar, Kumarawamy, and
  Kotiveerachari}]{Devaraju2013}
\bibinfo{author}{A.~Devaraju}, \bibinfo{author}{A.~Kumar},
  \bibinfo{author}{A.~Kumarawamy}, \bibinfo{author}{B.~Kotiveerachari},
\newblock \bibinfo{title}{Influence of reinforcements (sic and al2o3) and
  rotational speed on wear and mechanical properties of aluminum alloy 6061-t6
  based surface hybrid composites produced via friction stir processing},
\newblock \bibinfo{journal}{Materials and Design} \bibinfo{volume}{51}
  (\bibinfo{year}{2013}) \bibinfo{pages}{331--341}.
\bibitem[{Giovino et~al.(2018)Giovino, Pribyl, Benicewicz, and
  Bucinell}]{Giovino2018}
\bibinfo{author}{M.~Giovino}, \bibinfo{author}{J.~Pribyl},
  \bibinfo{author}{B.~Benicewicz}, \bibinfo{author}{R.~Bucinell},
\newblock \bibinfo{title}{Mechanical properties of polymer grafted nanoparticle
  composites},
\newblock \bibinfo{journal}{Nanocomposites} \bibinfo{volume}{4}
  (\bibinfo{year}{2018}) \bibinfo{pages}{244–252}.
\bibitem[{Thostenson and Chou(2002)}]{Thostenson2002}
\bibinfo{author}{E.~T. Thostenson}, \bibinfo{author}{T.-W. Chou},
\newblock \bibinfo{title}{Aligned multi-walled carbon nanotube-reinforced
  composites: processing and mechanical characterization},
\newblock \bibinfo{journal}{Journal of Physics D: Applied Physics}
  \bibinfo{volume}{35} (\bibinfo{year}{2002}) \bibinfo{pages}{L77--L80}.
\bibitem[{Gojny et~al.(2005)Gojny, Wichmann, Fiedler, and Schulte}]{Gojny2005}
\bibinfo{author}{F.~H. Gojny}, \bibinfo{author}{M.~H.~G. Wichmann},
  \bibinfo{author}{B.~Fiedler}, \bibinfo{author}{K.~Schulte},
\newblock \bibinfo{title}{Influence of different carbon nanotubes on the
  mechanical properties of epoxy matrix composites – a comparative study},
\newblock \bibinfo{journal}{Composites Science and Technology}
  \bibinfo{volume}{65} (\bibinfo{year}{2005}) \bibinfo{pages}{2300--2313}.
\bibitem[{Hári et~al.(2017)Hári, Horváth, Móczó, Renner, and
  Pukánszky}]{Hari2017}
\bibinfo{author}{J.~Hári}, \bibinfo{author}{F.~Horváth},
  \bibinfo{author}{J.~Móczó}, \bibinfo{author}{K.~Renner},
  \bibinfo{author}{B.~Pukánszky},
\newblock \bibinfo{title}{Competitive interactions, structure and properties in
  polymer/layered silicate nanocomposites},
\newblock \bibinfo{journal}{Polymer Letters} \bibinfo{volume}{11}
  (\bibinfo{year}{2017}) \bibinfo{pages}{479–492}.
\bibitem[{Weng et~al.(2016)Weng, Wang, Senthil, and Wu}]{Weng2016}
\bibinfo{author}{Z.~Weng}, \bibinfo{author}{J.~Wang},
  \bibinfo{author}{T.~Senthil}, \bibinfo{author}{L.~Wu},
\newblock \bibinfo{title}{Mechanical and thermal properties of
  abs/montmorillonite nanocomposites for fused deposition modeling 3d
  printing},
\newblock \bibinfo{journal}{Materials and Design} \bibinfo{volume}{102}
  (\bibinfo{year}{2016}) \bibinfo{pages}{276--283}.
\bibitem[{Abenojar et~al.(2017)Abenojar, Tutor, Ballesteros, del Real, and
  Martínez}]{Abenojar2017}
\bibinfo{author}{J.~Abenojar}, \bibinfo{author}{J.~Tutor},
  \bibinfo{author}{Y.~Ballesteros}, \bibinfo{author}{J.~C. del Real},
  \bibinfo{author}{M.~A. Martínez},
\newblock \bibinfo{title}{Erosion-wear, mechanical and thermal properties of
  silica filled epoxy nanocomposites},
\newblock \bibinfo{journal}{Composites Part B: Engineering}
  \bibinfo{volume}{120} (\bibinfo{year}{2017}) \bibinfo{pages}{42--53}.
\bibitem[{Hughes and Kelly(1953)}]{Hughes-Kelly1953}
\bibinfo{author}{D.~S. Hughes}, \bibinfo{author}{J.~L. Kelly},
\newblock \bibinfo{title}{Second-order elastic deformation of solids},
\newblock \bibinfo{journal}{Physical Review} \bibinfo{volume}{92}
  (\bibinfo{year}{1953}) \bibinfo{pages}{1145--1149}.
\bibitem[{Belashov et~al.(2020)Belashov, Beltukov, Moskalyuk, and
  Semenova}]{PolymTest2020}
\bibinfo{author}{A.~V. Belashov}, \bibinfo{author}{Y.~M. Beltukov},
  \bibinfo{author}{O.~A. Moskalyuk}, \bibinfo{author}{I.~V. Semenova},
\newblock \bibinfo{title}{Relative variations of nonlinear elastic moduli of
  polystyrene-based nanocomposites},
\newblock \bibinfo{journal}{Polymer Testing,} \bibinfo{volume}{submitted}
  (\bibinfo{year}{2020}).
\bibitem[{Samsonov(2001)}]{ams}
\bibinfo{author}{A.~Samsonov}, \bibinfo{title}{Strain Solitons in Solids and
  how to construct them}, \bibinfo{publisher}{Chapman \&Hall/CRC Press, Boca
  Raton, London, New York}, \bibinfo{year}{2001}.
\bibitem[{Samsonov et~al.(2017)Samsonov, Semenova, and Belashov}]{WaMot2017}
\bibinfo{author}{A.~M. Samsonov}, \bibinfo{author}{I.~V. Semenova},
  \bibinfo{author}{A.~V. Belashov},
\newblock \bibinfo{title}{Direct determination of bulk strain soliton
  parameters in solid polymeric waveguides},
\newblock \bibinfo{journal}{Wave Motion} \bibinfo{volume}{71}
  (\bibinfo{year}{2017}) \bibinfo{pages}{120--126}.
\bibitem[{Belashov et~al.(2018)Belashov, Beltukov, and Semenova}]{spie2018}
\bibinfo{author}{A.~V. Belashov}, \bibinfo{author}{Y.~M. Beltukov},
  \bibinfo{author}{I.~V. Semenova},
\newblock \bibinfo{title}{Pump-probe digital holography for monitoring of long
  bulk nonlinear strain waves in solid waveguides},
\newblock \bibinfo{journal}{Proceedings of SPIE} \bibinfo{volume}{10678}
  (\bibinfo{year}{2018}) \bibinfo{pages}{1067810--1--7}.
\bibitem[{Dreiden et~al.(2010)Dreiden, Khusnutdinova, Samsonov, and
  Semenova}]{strain2010}
\bibinfo{author}{G.~V. Dreiden}, \bibinfo{author}{K.~R. Khusnutdinova},
  \bibinfo{author}{A.~M. Samsonov}, \bibinfo{author}{I.~V. Semenova},
\newblock \bibinfo{title}{Longitudinal strain solitary wave in a two-layered
  polymeric bar},
\newblock \bibinfo{journal}{Strain} \bibinfo{volume}{46} (\bibinfo{year}{2010})
  \bibinfo{pages}{589–598}.
\bibitem[{Dreiden et~al.(2014)Dreiden, Samsonov, and Semenova}]{tpl2014}
\bibinfo{author}{G.~Dreiden}, \bibinfo{author}{A.~Samsonov},
  \bibinfo{author}{I.~Semenova},
\newblock \bibinfo{title}{Observation of bulk strain solitons in layered bars
  of different materials},
\newblock \bibinfo{journal}{Technical Physics Letters} \bibinfo{volume}{40}
  (\bibinfo{year}{2014}) \bibinfo{pages}{1140--1141}.
\bibitem[{Belashov et~al.(2018)Belashov, Beltukov, Petrov, Samsonov, and
  Semenova}]{apl2018}
\bibinfo{author}{A.~V. Belashov}, \bibinfo{author}{Y.~M. Beltukov},
  \bibinfo{author}{N.~V. Petrov}, \bibinfo{author}{A.~M. Samsonov},
  \bibinfo{author}{I.~V. Semenova},
\newblock \bibinfo{title}{Indirect assessment of bulk strain soliton velocity
  in opaque solids},
\newblock \bibinfo{journal}{Applied Physics Letters} \bibinfo{volume}{112}
  (\bibinfo{year}{2018}) \bibinfo{pages}{121903}.
\bibitem[{Dreiden et~al.(2008)Dreiden, Khusnutdinova, Samsonov, and
  Semenova}]{JAP2008}
\bibinfo{author}{G.~V. Dreiden}, \bibinfo{author}{K.~R. Khusnutdinova},
  \bibinfo{author}{A.~M. Samsonov}, \bibinfo{author}{I.~V. Semenova},
\newblock \bibinfo{title}{Comparison of the effect of cyanoacrylate- and
  polyurethane-based adhesives on a longitudinal strain solitary wave in
  layered polymethylmethacrylate waveguides},
\newblock \bibinfo{journal}{Journal of Applied Physics} \bibinfo{volume}{104}
  (\bibinfo{year}{2008}) \bibinfo{pages}{086106}.
\bibitem[{Nair et~al.(2000)Nair, Kumar, Thomas, Schit, and Ramamurthy}]{naira}
\bibinfo{author}{K.~Nair}, \bibinfo{author}{R.~Kumar},
  \bibinfo{author}{S.~Thomas}, \bibinfo{author}{S.~Schit},
  \bibinfo{author}{K.~Ramamurthy},
\newblock \bibinfo{title}{Rheological behavior of short sisal fiber-reinforced
  polystyrene composites},
\newblock \bibinfo{journal}{Composites Part A: Applied Science and
  Manufacturing} \bibinfo{volume}{31} (\bibinfo{year}{2000})
  \bibinfo{pages}{1231--1240}.
\bibitem[{Gelfer et~al.(2003)Gelfer, Song, Liu, Hsiao, Chu, Rafailovich, Si,
  and Zaitsev}]{gelfer}
\bibinfo{author}{M.~Y. Gelfer}, \bibinfo{author}{H.~H. Song},
  \bibinfo{author}{L.~Liu}, \bibinfo{author}{B.~S. Hsiao},
  \bibinfo{author}{B.~Chu}, \bibinfo{author}{M.~Rafailovich},
  \bibinfo{author}{M.~Si}, \bibinfo{author}{V.~Zaitsev},
\newblock \bibinfo{title}{Effects of organoclays on morphology and thermal and
  rheological properties of polystyrene and poly(methyl methacrylate) blends},
\newblock \bibinfo{journal}{Journal of Polymer Science Part B: Polymer Physics}
  \bibinfo{volume}{41} (\bibinfo{year}{2003}) \bibinfo{pages}{44--54}.
\bibitem[{Yadav et~al.(2020)Yadav, Chrysochoos, Arnould, and
  Bardet}]{Yadav-2020}
\bibinfo{author}{P.~Yadav}, \bibinfo{author}{A.~Chrysochoos},
  \bibinfo{author}{O.~Arnould}, \bibinfo{author}{S.~Bardet},
\newblock \bibinfo{title}{Effect of thermomechanical couplings on viscoelastic
  behaviour of polystyrene},
\newblock in: \bibinfo{booktitle}{Dynamic Behavior of Materials, Volume 1},
  \bibinfo{publisher}{Springer}, \bibinfo{year}{2020}, pp.
  \bibinfo{pages}{17--24}.

\end{thebibliography}







\end{document}